\documentclass[12pt]{article}

\usepackage{scicite}
\usepackage{graphicx}
\usepackage{times}
\usepackage{color}
\usepackage{amsmath,amssymb}

\newenvironment{sciabstract}{%
\begin{quote} \bf}
{\end{quote}}

\newcommand{\W}{{\mathbf W}_{F}}
\newcommand{\WL}{{\mathbf W}_{L}}

\renewcommand{\Re}{\operatorname{Re}}

\title{Optimal Network Topology for Responsive Collective Behavior}

\author
{David Mateo,$^{1\ast}$
Nikolaj Horsevad,$^{1}$
Vahid Hassani,$^{1}$ \\
Mohammadreza Chamanbaz$^{1,2}$,
Roland Bouffanais$^{1}$\\
\\
\normalsize{$^{1}$ Singapore University of Technology and Design,}\\
\normalsize{8 Somapah Road, Singapore 487372}\\
\normalsize{$^{2}$ Arak University of Technology, Daneshgah Road, Arak, Iran}\\
\\
\normalsize{$^\ast$To whom correspondence should be addressed; E-mail: david.mateo.valderrama@gmail.com.}
}

\date{}

\begin{document}


\baselineskip24pt


\maketitle


\begin{sciabstract}
    Animals, humans, and robotic multi-agent systems usually operate in dynamic environments, where the ability to respond to changing circumstances is of paramount importance.
    An effective collective response requires suitable information transfer among agents, and thus is critically dependent on the agents' interaction network.
    In order to investigate the influence of the network topology on
    collective response, we consider an archetypal model of distributed
    decision-making---the leader-follower linear consensus---and study the
    collective capacity of the system to follow a dynamic driving signal (the
    ``leader'', an external perturbation locally disturbing the
    collective dynamics) for a range of topologies and system sizes.
    Experiments on the collective behavior of a swarm of land robots reveal a nontrivial relationship between the frequency of the driving signal and the optimal network topology.
    The emergent collective response of the swarm to a slow-changing perturbation increases with the degree of the interaction network, but the opposite is true for the response to a fast-changing one.
    In general, the collective response in consensus dynamics is optimal when each individual interacts with a certain number of agents that decreases monotonically with the frequency and, for large enough systems, is independent of the size of the system.
    These results have far-reaching practical implications for the design and understanding of distributed systems, since they highlight that a dynamic rewiring of the interaction network is essential to the effective collective operations of multi-agent systems at different time-scales.
\end{sciabstract}


\section*{Introduction}\label{sec:introduction}

A wide range of complex systems are characterized by relatively simple
dynamical rules, while still producing excessively complex emergent collective
behaviors. Examples abound in the natural world (e.g., flock of birds, school
of fish, swarm of insects~\cite{%
  krause02:_livin_in_group,bouffanais15:_desig_contr_swarm_dynam,cavagna18,young13:_starl_flock_networ_manag_uncer,attanasi14:_collec_behav_collec_order_wild_swarm_midges,viscek2012,swain2015coordinated,Mat17,shang14:_consen}), in social
systems (e.g., social networks~\cite{%
  fowler10:_cooper,centola10:_spread_behav_onlin_social_networ_exper,amelkin2017polar}), as
well as in engineered multi-agent systems (e.g., self-organized networks of
mobile sensors, multi-vehicle coordination, and swarm robotics systems~\cite{%
  ren08:_distr,%
  chamanbaz17:_swarm_enabl_techn_multi_robot_system,%
  zoss17:_distr_system_auton_buoys_scalab,lin14}).

Historically, significant attention has been directed towards investigating
varieties of collective behaviors obtained by testing a wide range of local agent-to-agent
interaction rules~\cite{viscek2012,shang14:_consen}. Collective behaviors have also been investigated from the
network-theoretic
perspective~\cite{komareji13:_resil_contr_dynam_collec_behav,shang14:_influen,punzo16:_using,liu16:_contr,young13:_starl_flock_networ_manag_uncer,komareji18:_consen,Mat17}. It
is now clear that such rich collective behaviors are the outcome of a complex
interplay between network topology---characteristic of the group-level
organization---and the dynamical laws at the agent's
level~\cite{sekunda16:_inter,liu16:_contr,young13:_starl_flock_networ_manag_uncer,komareji18:_consen,Mat17}.

Many collective behaviors can be studied through the lens of distributed consensus problems,
including collective motion in animal groups and multi-robot systems.
Over the past decade, the number of studies on decentralized consensus and cooperation in networked
multi-agent systems has experienced a spectacular growth, with concomitant
developments in various fields of engineering and
science~\cite{cao2013overview,olfati-saber07:_consen_cooper_networ_multi_agent_system,cavagna18,bouffanais15:_desig_contr_swarm_dynam,pirani2016smallest,clark2012leader}. Consensus
dynamics is the cornerstone of cooperative control strategies for vehicular
formation~\cite{ren08:_distr,cao2013overview,lin14}, swarm robotics~\cite{chamanbaz17:_swarm_enabl_techn_multi_robot_system,%
  zoss17:_distr_system_auton_buoys_scalab}, and synchronization of coupled
oscillators~\cite{li2010consensus,cao2013overview}. Decentralized
consensus is also at the core of collective opinion dynamics and complex contagion processes
in social
networks~\cite{fowler10:_cooper,centola10:_spread_behav_onlin_social_networ_exper,amelkin2017polar},
as well as complex collective responses in biological swarms~\cite{attanasi14:_collec_behav_collec_order_wild_swarm_midges,viscek2012,cavagna18,young13:_starl_flock_networ_manag_uncer,swain2015coordinated,Mat17}.

Past studies focused on establishing the influence of the interaction network
topology on (i) the capacity of the collective to reach consensus in the presence
of noise, communication constraints, and time
delays~\cite{komareji18:_consen,cao2013overview}, (ii)
the speed of consensus (i.e. its convergence
rate)~\cite{Alm07,shang14:_influen,pirani2016smallest}, (iii) the
stability and stabilization of consensus~\cite{cao2013overview}, and (iv) the ability to steer the system
toward a particular consensus value by means of various control techniques such as
pinning control, cooperative tracking control or model reference consensus~\cite{liu16:_contr,punzo16:_using}.
However, the effects of the network topology on other dynamical properties of
distributed multi-agent systems such as their adaptivity or responsiveness to
external perturbations have received considerably less
attention~\cite{young13:_starl_flock_networ_manag_uncer}.

It is important stressing that a capacity for fast consensus is not necessarily indicative of a responsive collective behavior.
For instance, ferromagnets at low temperature exhibit a global spontaneous magnetization---a process that can be described by a distributed consensus protocol.
It is known that both the degree of consensus (i.e. magnetization) and the speed at which it is reached increase with decreasing temperature,
but the capacity of the system to respond to external perturbations is maximized at a finite critical temperature.

Similarly, in the context of animal collective motion it has been observed that midges exhibit low levels of ordering while maintaining large connected correlations, thus having a high collective response~\cite{attanasi14:_collec_behav_collec_order_wild_swarm_midges}.
With these observations, the authors eloquently argued that
one must be careful in relating collective order (i.e. degree of consensus) with the collective responsiveness.
The latter was obtained experimentally by measuring the correlations in the fluctuations of their behavior.
While inferring a collective response to external perturbations from these fluctuations is not formally justified for out-of-equilibrium systems, extensive numerical studies~\cite{Cha08} have shown that this equivalence holds in the context of collective motion based on distributed heading consensus.
Moreover, simulations have shown that this measure of susceptibility is a good
indicator of the group's performance in biologically-relevant functions such
as predator avoidance~\cite{Mat17}. These facts along with other empirical
evidence have led to the conclusion that responsiveness, rather than high
consensus or order, is the true hallmark of collective behavior~\cite{cavagna18}.

To study how the responsiveness of a collective is affected by its interaction network topology, we consider an elementary example of distributed decision-making: a linear time-invariant (LTI) system of agents performing consensus over a scalar state-variable.
The agents---i.e., the nodes of the interaction network---are all identical,
except for one ``leader'' (also known as ``stubborn'' agent in some
contexts~\cite{pirani2016smallest,amelkin2017polar}) with some arbitrary pre-defined
dynamics. From the control-theoretic perspective, this leader introduces a
time-varying control input signal into the system. In the biological context,
this dynamical leader represents a member of a swarm with access to
privileged information about a food source or a threat, i.e. the dynamics of
the leader can be considered to be a reaction to some external perturbation within
the environment. For instance, in a school of fish, a single fish detecting
the approach of a predator swiftly changes its direction of travel, and this
rapid signal---the local external perturbation---propagates through the school
thereby triggering a large-scale
evasive maneuver whose effectiveness is critical to the
survival of the group.

Investigating the propagation of a local perturbation with a possibly
broad spectrum of time-scales is of critical importance to a vast breadth of
decentralized networked systems: e.g. the fast shutdown at one
end of a power grid can cascade into a large-scale blackout, a snowstorm at
one critical node of an airport network generates delays throughout the entire
system, and a fad introduced
by an ``influencer'' propagates
and amplifies through a scale-free social network.
For artificial multi-agent systems, introducing a leader can facilitate a range of formation control techniques~\cite{ren08:_distr, lin14}
by means of pinning control or cooperative tracking
control~\cite{liu16:_contr,punzo16:_using,clark2012leader}.
In such a scenario, having a responsive collective behavior is crucial in the
case that the target formation changes with time.

Here, the leader provides a gateway to injecting local perturbations in the emergent collective behavior.
This allows us to study the ensuing collective response
as a function of the time-scale of the injected perturbation and the topology of the interaction network.
A similar LTI framework was considered in~\cite{young13:_starl_flock_networ_manag_uncer}
to study the effect of the network
topology on the
consensus.
However, the study was limited to the long-time consensus dynamics arising from a global perturbation---in the form of white noise injected into the
dynamics of all agents.

To quantify the collective responsiveness, several metrics are used
across different communities---susceptibility, gain, or frequency
response---but the definition is consistent: the response is measured by the
rate of change in the variable of interest (in our case, the state variable
participating in the consensus dynamics) with respect to a change in the
external perturbation applied (in our case, the input signal injected through
the leader).
For a linear model, this rate of change can be expressed analytically as
Eq.~\ref{eq:gain}, and is a complex $N$-vector in the state-space of agents (see Methods).
We can quantify the {\it collective} frequency response of the system using
the square of the norm of this vector $H^2$.
Since each component $h_i(\omega)$ of $\mathbf{H}(\omega)$ corresponds to the
individual frequency response of a participating agent $i$, and given that $|h_i|\leq 1$, the collective frequency response
$H^2(\omega)$ can be interpreted as the number of agents that are able
to respond or follow the leader, when the dynamics of the latter varies with frequency $\omega$.
From a statistical perspective, $H^2$ measures the size of the fluctuations in the consensus state-variable induced by a localized perturbation.
For a connected topology, one expects $H^2\simeq N$ at low frequency, i.e. for
frequencies below the system's speed of consensus as determined by the smallest eigenvalue of the grounded
Laplacian~\cite{pirani2016smallest}. In the limit of zero frequency, the
system is at steady state and the consensus is guaranteed, therefore $H^2(\omega=0)=N$.
At higher frequency, the collective response always decays with increasing
$\omega$, but the way that it decays is intricately dependent on the details
of the connectivity between agents.

We show that the degree distribution of the interaction network is a
central element in controlling the responsiveness of the collective.
Even a relatively unadorned model, such as the linear consensus protocol, displays a rich phenomenology that crucially depends on the time-scale of the changes in the input signal. Specifically, when the system
is driven at low frequency, an increase in the number of interagent
connections always improves the collective response.
At high frequency, on the
contrary, reducing the agents' degree benefits the responsiveness. More
generally, in the presence of a dynamic driving signal changing at a given time-scale, it is
possible to tune the interaction network in
order to maximize the collective response.
Lastly, we present
experimental validations of our findings with a swarm of 11 robots performing heading consensus.
We measure the capacity of the robots to align their orientation to a ``leader'' agent as a function of the degree distribution.
From low to high frequency, the variations of the measured collective response with the degree of the interaction network are in very good agreement with the trend predicted by the distributed linear consensus theory.

These findings have far-reaching implications for the design of artificial swarms or interaction
networks.
They teach us that, in order to maximize the collective response, the system requires the ability to dynamically adjust the degree of the interaction network depending on the time-scale at which it should respond.

\section*{Results}

\subsection*{Influence of the number of connections on the frequency response}\label{results:number_of_neis}

In Ref.~\cite{Mat17}, it was shown that a breadth of networked systems can maximize their performance by tuning the number of connections to a specific finite value: a self-propelled particle swarm can increase its capacity to avoid a predator, and a collective performing distributed decision-making can react to an external influence faster.
In the case of distributed consensus, it was seen that the collective frequency response of the system to a single leader when the agents are connected by a regular one-dimensional grid (a ring) is maximized for a number of neighbors $k^*$ that depends on the frequency of the leader.

\begin{figure}
    \includegraphics{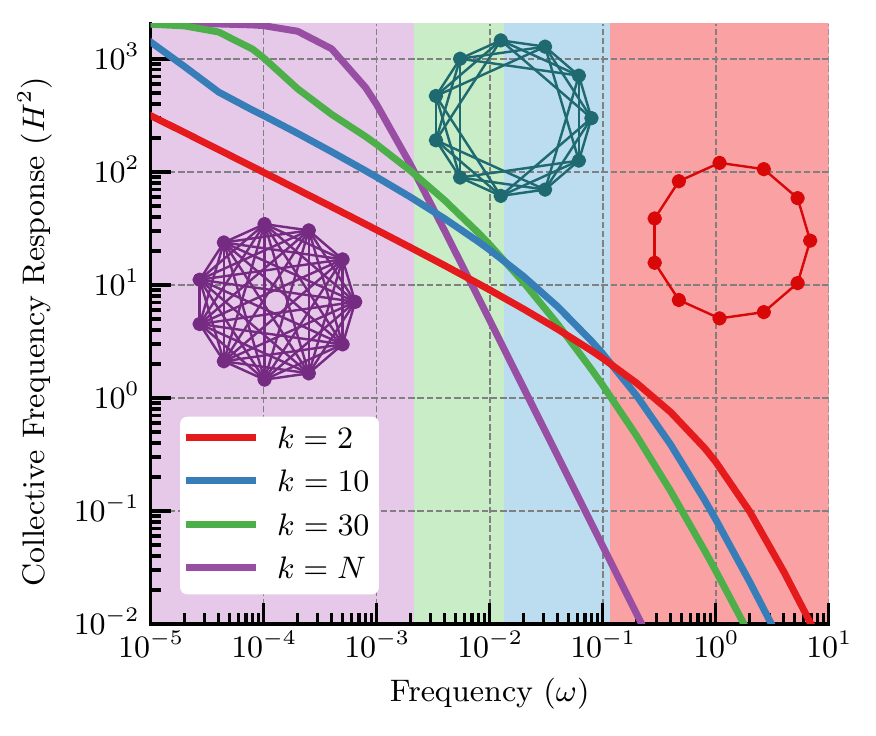}
        \includegraphics{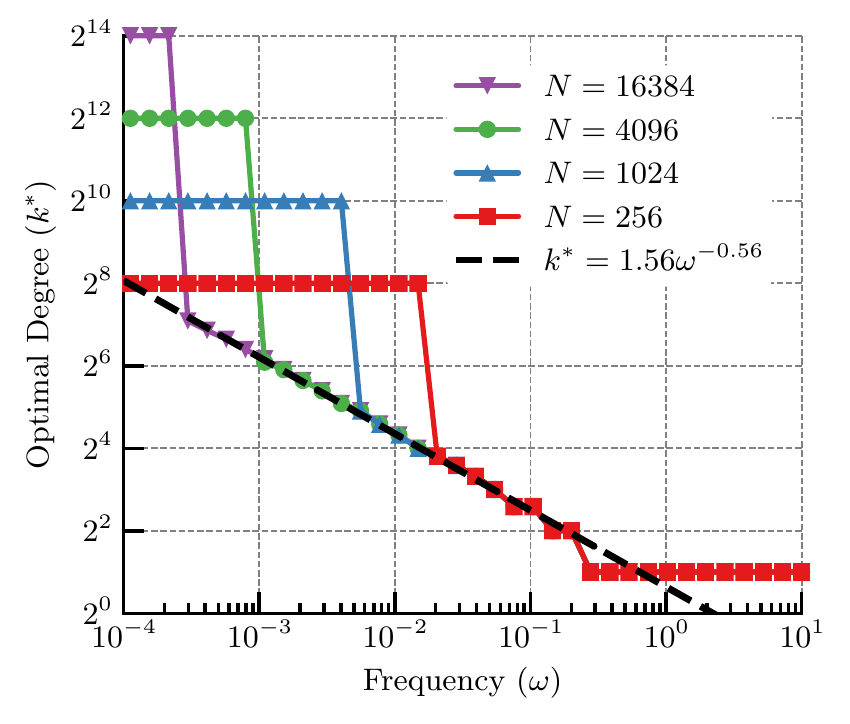}
        \caption{
        {\it Collective frequency response for a ring network.}
        {\bf Left:}
        Response of $N=2048$ agents performing distributed linear consensus over a regular periodic one-dimensional grid with fixed degree $k$.
        Larger degrees yield a higher response at low frequency ($\omega < 2\times10^{-3}$) while the opposite is true at high frequency ($\omega > \omega_{\textrm{high}} = 0.278$).
        {\bf Right:}
        Optimal degree $k^*$ for maximum collective response as a function of the frequency $\omega$ for a system of $N$ agents distributed on a ring.
        For low frequency, the optimal $k^*$ corresponds to an all-to-all connectivity.
        At higher frequencies, $k^*(\omega)$ follows the ``bulk'' behavior from Eq.~\ref{eq:kstar} (fit, black line) up to its lowest possible value, $k^* = 2$, at $\omega=\omega_{\textrm{high}}$.
        }
        \label{fig:ring}
\end{figure}

Figure~\ref{fig:ring} shows the collective frequency response of a system performing distributed linear consensus (Eq.~\ref{eq:dynamics}) when one leader and $N=2048$ agents are connected to their $k$-closest neighbors in a ring topology.
When all agents are connected to each other ($k=N$), the response of the system is a first-order low-pass filter with a cutoff frequency of $1/N$.
This all-to-all connectivity provides an optimal response for any frequency below $\omega_{\textrm{low}} \simeq 2\times10^{-3}$.
Above this threshold, a lower network degree yields a higher response.

The higher the frequency, the better low degrees (small $k$) respond as compared to high (large $k$) ones.
For instance, $k=30$ yields a higher response than $k=N$ for $\omega \ge 2.24\times10^{-3}$.
In turn, the response of the system with $k=10$ exceeds that of $k=30$ for frequencies $\omega \ge 1.38\times10^{-2}$.
A minimal, closest-neighbor connectivity ($k=2$) provides the optimal response for any frequency above $\omega_{\textrm{high}} \simeq 2.78\times10^{-1}$.

The optimal degree $k^*$ corresponding to the highest response at a given frequency $\omega$ is displayed in Fig.~\ref{fig:ring} for several system sizes $N$.
For large enough systems, the optimal degree follows a scaling law of the form
\begin{equation}
    k^*(\omega) = K_0 \omega^{-\gamma} ,
    \label{eq:kstar}
\end{equation}
with $K_0=1.56$ and $\gamma=0.56$.
Interestingly, the optimal degree does not scale with the size of the system $N$ in any observable way.
The finite-size effects manifest at low frequency, where $k^*$ ``jumps'' from its bulk value given by Eq.~\ref{eq:kstar} to $k^* = N$.


The frequency at which this jump occurs does depend on the size of the system, and corresponds approximately to $k^* = \sqrt{N}$.
In the limit $N\rightarrow \infty$, a finite connectivity is always preferable over the all-to-all topology for any finite frequency.

\begin{figure}
    \includegraphics{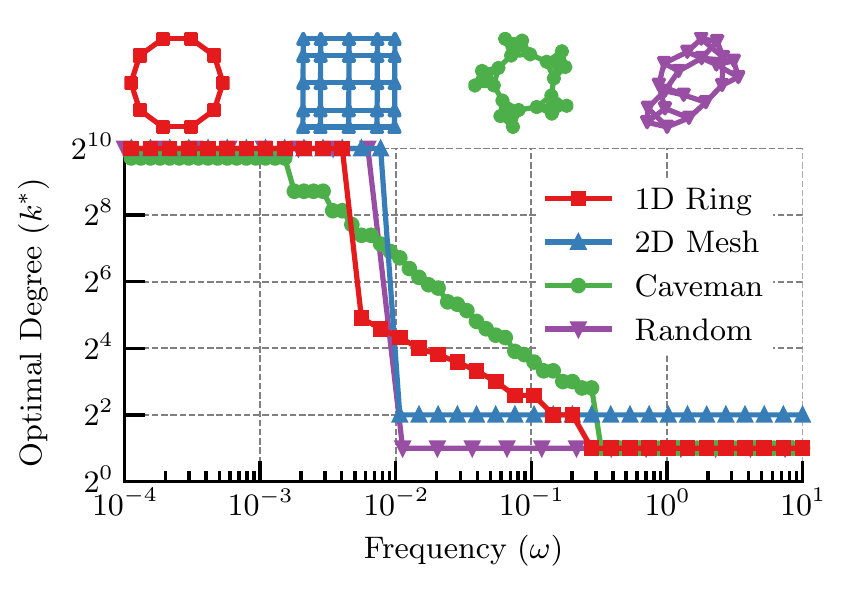}\includegraphics{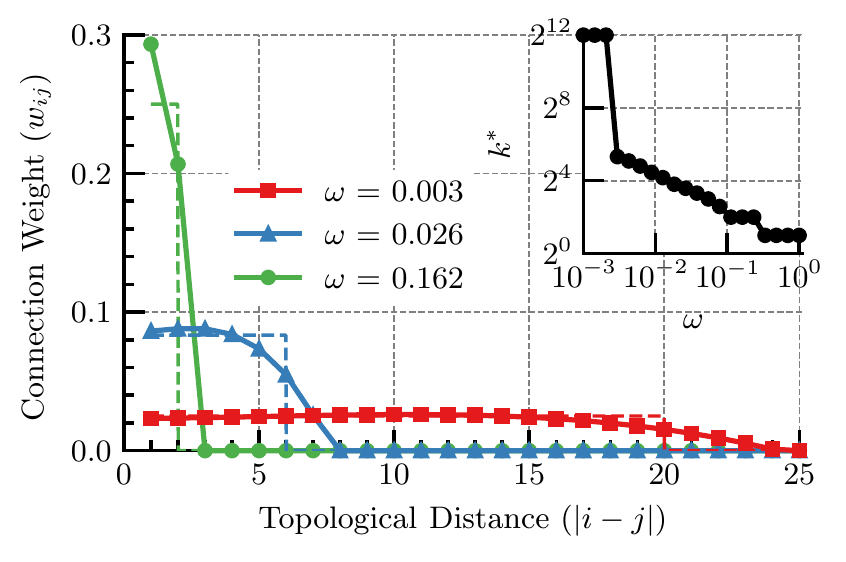}
    \includegraphics{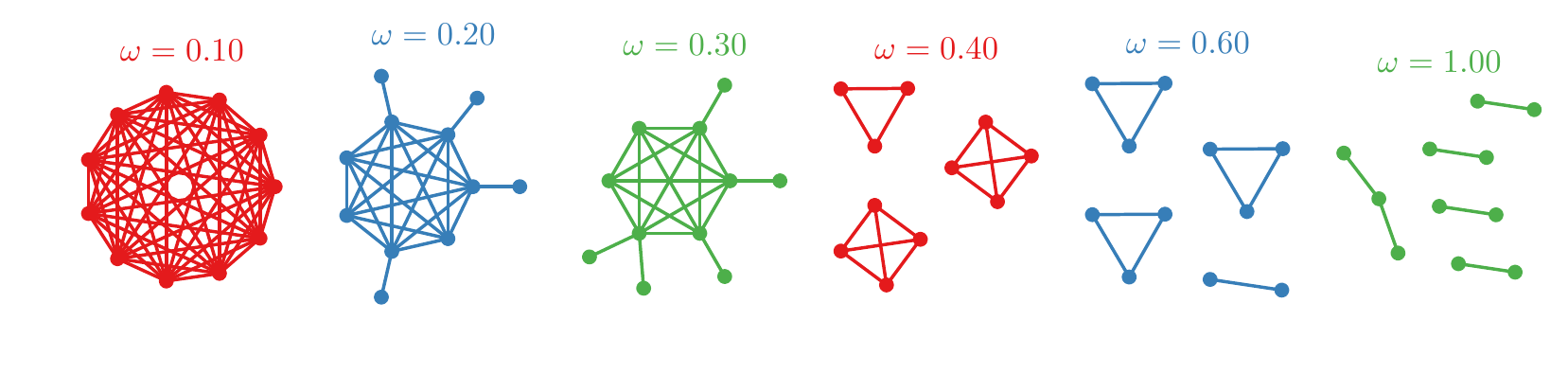}
    \caption{{\it Optimal networks for collective response.}
    {\bf Top left:}
    Optimal degree $k^*$ for maximum collective response as a function of the frequency $\omega$ for a system of $N=1024$ arranged in different network topologies
    ($N=840$ for the caveman topology).
    Note that some networks display a sudden transition from all-to-all to minimal connectivity, while others have an intermediate range of frequencies where $k^*$ follows a scaling law of the form Eq.~\ref{eq:kstar}.
    {\bf Top right:}
    Optimal connection weights $w_{ij}$ as a function of topological distance for a system of $N=4096$ agents at a given frequency $\omega$.
    The inset shows the optimal number of connections $k^*$ obtained by fitting a Heaviside function to the weight distributions.
    {\bf Bottom:}
    Optimal network topologies for a system of $N+1=11$ agents obtained by stochastic numerical discrete optimization over the space of unweighted, undirected graphs.
    Instead of fixing the leader to be a particular agent, the optimization maximizes the collective frequency response averaged over all the possible leaders (allowing for disconnected graphs to be optimal).
    Note that the mean degree of these networks is consistently reduced with increased frequency $\omega$.}
    \label{fig:optimal}
\end{figure}

In general, the optimal degree for a given frequency depends on the network model considered.
Figure~\ref{fig:optimal} presents a comparison of the $k^*$ values obtained for different kinds of networks.
All networks considered display a $k^*$ that decreases monotonically with frequency, transitioning from an all-to-all connectivity at low frequency to a minimal connectivity at high ones.
Note that the minimum possible degree is $k=2$ for all models except for the 2D mesh that requires a minimum of $k=4$.
This transition from all-to-all to minimal connectivity may be abrupt (2D mesh, random), continuous (caveman model, where $k^*\simeq 1/\omega$), or a combination of both (1D ring).

%
%

\subsection*{Network optimization: weights and structure}\label{results:optimization}
The results in the previous section present a clear phenomenology for several types of networks that hints towards a possible universal behavior.
In order to explore how general this phenomenology is, we consider two cases for which we relax some of the assumptions made previously.
In the first case, we consider weighted networks, for which the connection between agents is not binary.
In the second case, we consider a general network optimization problem for a small system where no particular structure is imposed.

\subsubsection*{Optimal Weights}

Unweighted graphs, i.e.~networks where $a_{ij} = \{0, 1\}$, represent only a small subset of the possible connectivities between agents, and there is a priori no reason to assume that the optimal connectivity can be represented by an unweighted graph.
Thus, expanding the study to weighted graphs opens the possibility of finding networks with arbitrary distribution of weights that have a higher frequency response.

Using a linear parametrization of $\mathbf{W}$ (see Methods) for a regular ring, where $w_{ij}$ depends only on the topological distance between agents $d_{ij} = |i-j|$, the numerical optimization of Eq.~\ref{eq:cost} yields the connectivity profiles shown in Fig.~\ref{fig:optimal}.
For a ring of $N=4096$ agents, the optimal responsiveness at frequencies $\omega < 5\times 10^{-3}$ corresponds to a simple all-to-all connectivity (all $w_{ij} = 1/N$).
At higher frequencies, one obtains a smooth profile where the optimal connectivity decreases with distance.
Still, most nodes are either connected ($w_{ij}\simeq 1/k^*$) or disconnected ($w_{ij}\simeq 0$), with only a few near the transition having intermediate weights.
This profile is very similar to the unweighted case, and the effective number of neighbors inferred from the profile (see inset of Fig.~\ref{fig:optimal}) is identical to the values presented in the previous section.


This numerical optimization corroborates that allowing for weighted connections does not sensibly change the phenomenology observed, where limiting the number of connections to a certain frequency-dependent value optimizes the response of the system.

\subsubsection*{Optimal Structure}\label{sec:sa}

So far, we have considered particular network models for which the degree $k$ can be controlled.
All these models exhibit an optimal frequency response when the degree is set to a certain $k^*$ that decreases with increasing frequency.
However, the particular value of $k^*$ depends on the model.
The results do not guarantee that, in general, any network with a given fixed degree $k^*$ will yield a higher collective response than any other network with a different degree.

Does the relationship between the number of neighbors and frequency response still hold when considering arbitrary network structures?
Preliminary results of numerical discrete optimization for small systems suggest that it is indeed the case.
The bottom row of Fig.~\ref{fig:optimal} shows examples of the networks obtained by performing simulated annealing optimization for the frequency response of a system of $N+1=11$ agents for six frequencies $\omega$.
%

At frequencies $\omega \le 0.1$, the optimization procedure consistently yields an all-to-all connectivity.
For higher frequencies $\omega = 0.2$ and $0.3$, the stochastic nature of the optimization procedure generates slightly different configurations, but all of them contain a clique of 7 agents at $\omega=0.2$ (5 to 6 at $\omega=0.3$) with the rest of the agents in the periphery of the clique.
These configurations have an average degree $\langle k\rangle = 4.7$ for $\omega=0.2$ and $\langle k\rangle=3.6$ for $0.3$.
Further increasing the frequency yields disconnected graphs where the agents are distributed in clusters of either four, three, or two agents for $\omega = 0.4, 0.6$, and $1$ respectively.

While the current results are limited to small systems, they show that the existence of an optimal mean degree that decreases with increasing frequency is not a particular feature of the network models considered here, but a general phenomenology of systems performing linear distributed consensus.

\subsection*{Application to Swarm Robotics}

To showcase the application of these results to the design of interaction networks in the nascent field of swarm robotics, we perform a series of experiments on heading consensus with a swarm of land robots where each one aligns its direction of motion with that of their neighbors.
This form of consensus, inspired by Vicsek's model of collective motion in natural swarms~\cite{vicsek95:_novel}, is closely related to the first-order distributed linear consensus discussed previously.
However, an empirical implementation involves significant deviations from the ideal scenario.

For instance, while the dynamics of the robots are ultimately governed by physical processes that are continuous in time, the robots sense each other's state using asynchronous, discrete communications with stochastic delays.
The consensus protocol itself, like in the Vicsek model, is nonlinear owing to the state variable being an angular quantity as opposed to a scalar one.


The swarm is composed of 11 robots (see Methods) equipped with a custom ``swarm-enabling'' unit---providing data-processing power and distributed communications---that allows the robots to perform complex, decentralized cooperative control algorithms~\cite{chamanbaz17:_swarm_enabl_techn_multi_robot_system}.
Each robot sends periodically its heading to its $k$ neighbors.
One of the robots, the leader, is set to continuously rotate at a fixed frequency $\omega$ in the range $0.01 \, \mathrm{Hz} < \omega < 0.1 \, \mathrm{Hz}$.
The rest of $N=10$ robots are set to perform the heading consensus algorithm of Eq.~\ref{eq:heading_consensus}.
The experiments are run for at least four periods of leader rotation on an open space of around 20 m$^2$.
The experimental setup does not impose any significant restriction on the location, other than a relatively flat and regular ground and enough space for the robots to move freely.

\begin{figure*}
    \includegraphics{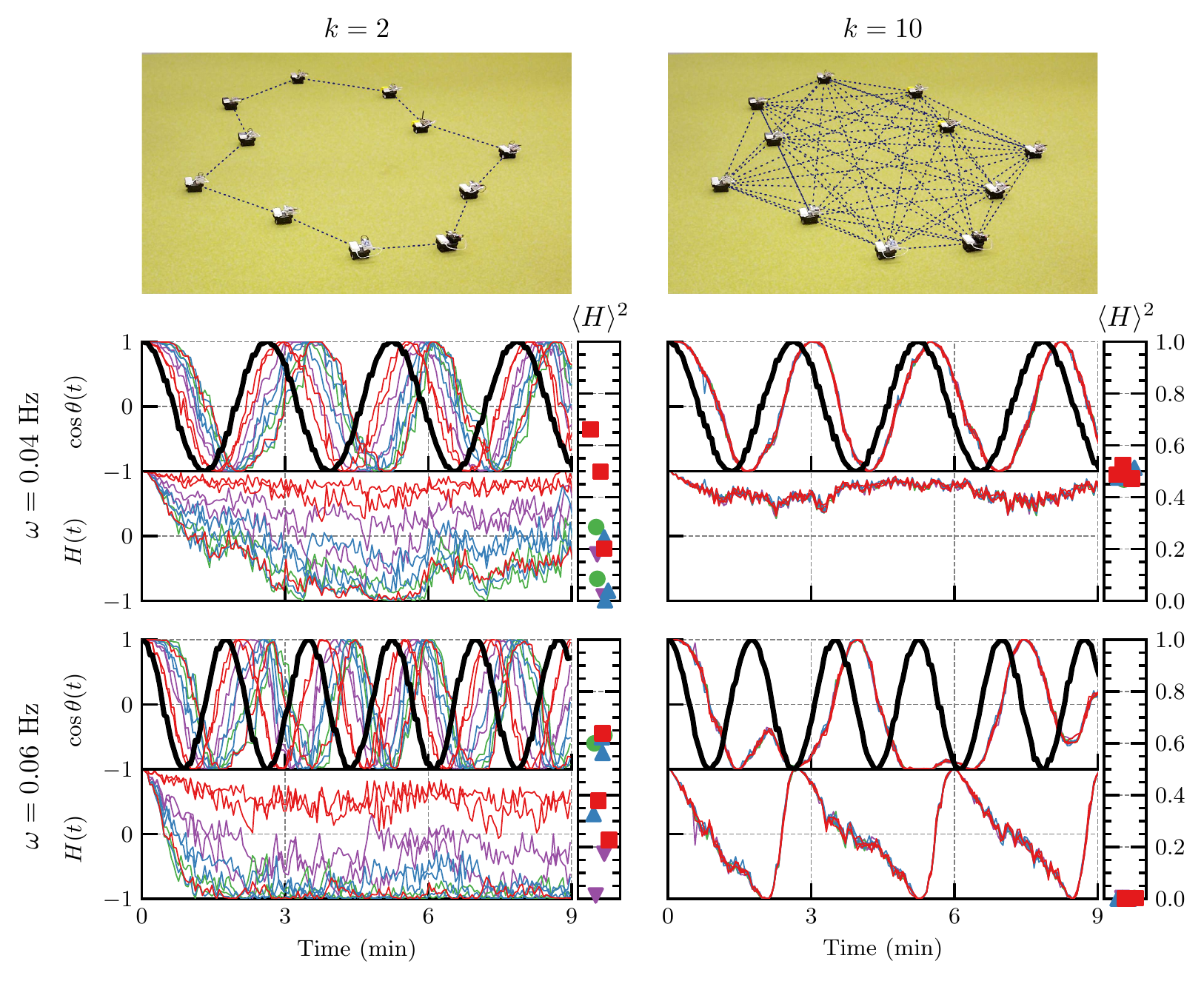}
    \caption{Evolution of the heading of 11 robots obtained in an experiment on heading consensus with one leader (black line) rotating at frequency $\omega = 0.04$~Hz (top) or $\omega = 0.06$~Hz (bottom) when each robot has either $k=2$ neighbors (left column) or $k=10$ (right column). The degree by which each agent is following the leader at a given instant is $H_i(t)$ (Eq.~\ref{eq:experimental_performance}) and the square of the mean value (displayed in the lateral bar) is its frequency response.
    }
    \label{fig:experiment_summary}
\end{figure*}

The effect of connectivity on responsiveness is investigated by repeating the experiment for different values of the number of neighbors $k$ and different rotation frequencies $\omega$ for the leader.
See Supplemental Material for videos of this experiment performed at low and high frequencies.
A characteristic selection of the results obtained is presented in Fig.~\ref{fig:experiment_summary}.
When the robots are connected by means of a ring topology ($k=2$, left column),
the qualitative behavior of the swarm is not critically affected by the frequency: both at $\omega=0.04$ Hz and $0.06$ Hz, the four or five robots topologically closest to the leader are able to reasonably follow it, with the rest lagging behind.

However, the picture changes when an all-to-all connectivity underpins the operations of the swarm ($k=10$, right column).
Since each robot has access to the same information, they all behave identically.
This allows the whole swarm to follow closely the leader at frequencies below a threshold of $\omega_c \simeq 0.05$~Hz,
but also causes the response of the system to drop drastically above this threshold.
In the context of the Vicsek model, the swarm has a low polarization when $k=2$ ($0.62$--$0.77$) and a high one when $k=10$ ($0.91$--$0.95$).
Here, we see that a high polarization allows the system to have a large coordinated response at low frequency, but it also tends to ``ossify'' the collective thus drastically reducing its response at high frequency.

This phenomenology is reminiscent of what has been observed in natural systems.
Specifically, it is known that while flocks of starlings possess high levels of ordering or polarization~\cite{attanasi14:_infor}, swarms of midges display low levels of polarization but a large connected correlation~\cite{attanasi14:_collec_behav_collec_order_wild_swarm_midges}, and thus high responsiveness to biologically-relevant environmental perturbations~\cite{Mat17}.

\begin{figure}
    \begin{minipage}[l]{0.6\linewidth}
        \includegraphics{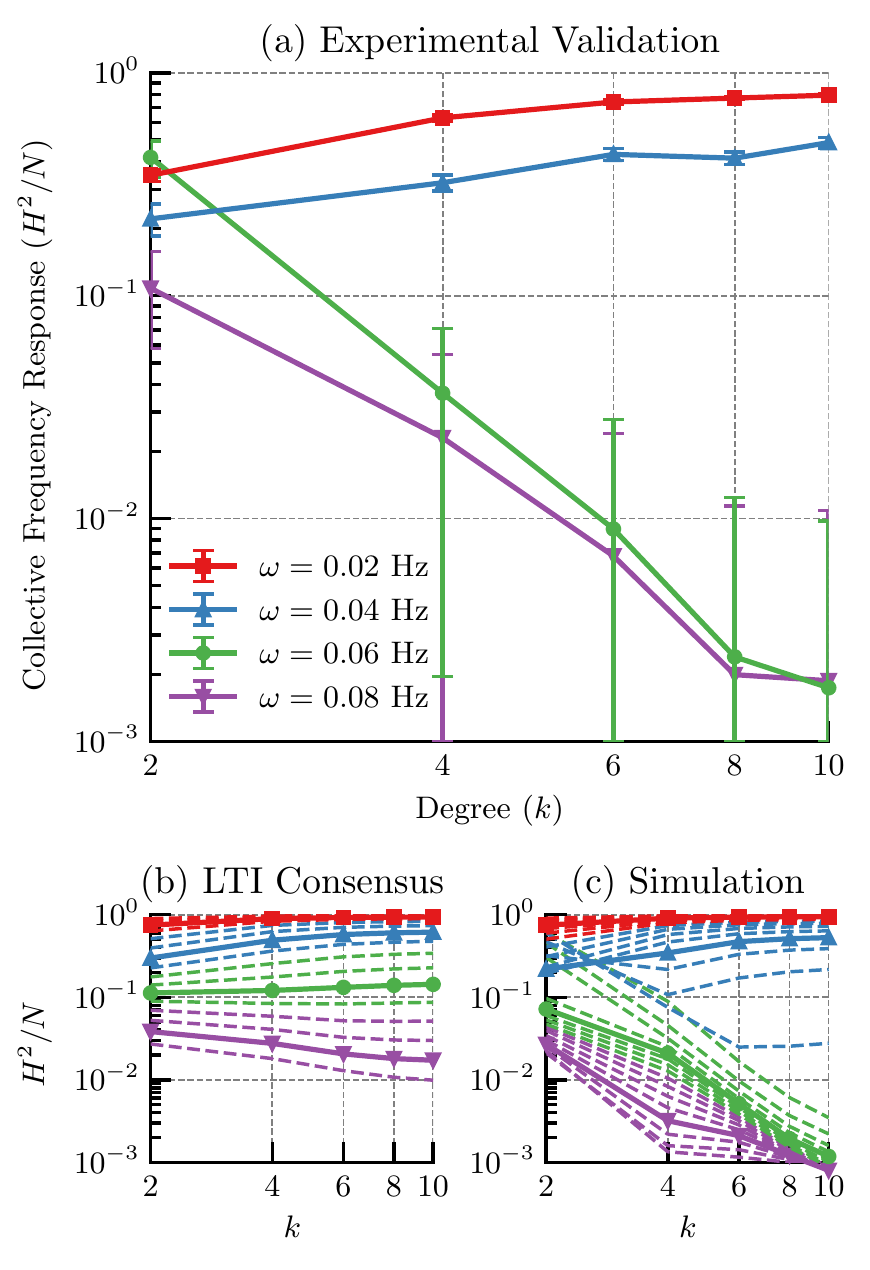}
    \end{minipage}
    \begin{minipage}[r]{0.4\linewidth}
        \caption{Collective response in leader-follower heading consensus for a system of $N+1=11$ agents.
        ({\bf a}) Experimental collective frequency response (Eq.~\ref{eq:experimental_gain}, normalized) obtained with 10
        robots performing distributed heading consensus plus one leader rotating at a fixed frequency $\omega$.
        ({\bf b}) Response for the equivalent LTI distributed linear consensus (Eq.~\ref{eq:gain}, normalized) with the same $N$.
        ({\bf c}) Response obtained with simulations of the heading consensus algorithm (Eq.~\ref{eq:heading_consensus}).}
        \label{fig:experiment_gain}
    \end{minipage}
\end{figure}

The measured capacity of the swarm to align to the time-dependent orientation of a leader is presented in Fig.~\ref{fig:experiment_gain} as a function of the number of neighbors for a selected number of frequencies.
These results confirm that the predictions from the distributed linear consensus model applied to complex, realistic cases of collective motion.
Specifically, the robots benefit from having more connections at low frequency but it hinders their responsiveness at high ones.

Unfortunately, our experimental system is too small to observe an optimal connectivity $k^*$ different than two or $N$, or to be able to meaningfully study different network models.
However, with the advent of large swarm robotics systems~\cite{rubenstein14:_progr},
it is likely that different values of $k^*$ can be experimentally measured for various network topologies.

\section*{Discussion}

This paper presents a detailed study on the relation between the collective frequency response to a time-dependent, localized signal (a leader) in a multi-agent system performing distributed consensus and the connectivity between the agents.
The study shows that the response to the driving signal arising from different connectivities depends critically on the time-scale of the signal: the smaller the time-scale, the better low-degree connectivities perform as compared with high-degree ones.

In particular, we observe that when the agents are connected by means of a static $k$-nearest neighbor ring configuration, the collective response of the system---excluding finite-size effects at low frequency---is maximized for a particular frequency-dependent number of neighbors that follows a scaling law of the form $k^* \propto \omega^{-0.56}$.
This functional form for $k^*(\omega)$ is not universal, but all the network models considered here consistently display a $k^*$ monotonically decreasing with the signal frequency $\omega$. For instance, a caveman network has a $k^* \propto \omega^{-1}$, and other networks such as a bidimensional grid or a regular random network have a bimodal $k^*$ that corresponds to an all-to-all connectivity for $\omega < \omega_c$ and to a minimal connectivity for $\omega > \omega_c$.

We posit that the existence of this optimal connectivity $k^*$, dependent on the time-scale $\tau=1/\omega$ of the local perturbation, has eluded prior investigations on distributed consensus because those did not consider either the localized nature of perturbations~\cite{komareji18:_consen,cao2013overview,shang14:_influen,young13:_starl_flock_networ_manag_uncer,vicsek95:_novel,viscek2012,Cha08, lin14} or the short-time transient component of the response~\cite{young13:_starl_flock_networ_manag_uncer,swain2015coordinated,ren08:_distr,clark2012leader, lin14}.
On the one hand,
the local injection of the perturbation into the collective dynamics is key to the uncovered phenomenology because it leads to a complex propagation of the input signal through a host of feedback and feedforward loops associated with the topology of the network.
On the other hand,
the short-time (high frequency) transient responsive behavior is not captured if only steady-state deviations from consensus, integrated metrics such as consensus speed, or similar proxies are used to characterize the response.

This phenomenology stands in stark contrast with that of the well-studied consensus-reaching process:
the prescription for a fast consensus is not necessarily compatible with the one for a responsive consensus.
The speed of consensus---a.k.a. convergence rate---can readily be obtained
from the second-smallest eigenvalue $\lambda_2$ of the Laplacian matrix  in the absence of
a
leader~\cite{olfati-saber07:_consen_cooper_networ_multi_agent_system},
or from the smallest eigenvalue $\lambda_1^g$ of the grounded Laplacian matrix in the
presence of a leader~\cite{pirani2016smallest}. In general, $\lambda_2$ and
$\lambda_1^g$ tend to increase with increasing mean
degree~\cite{Alm07,shang14:_influen,pirani2016smallest}. In particular, the
extreme case of an all-to-all connectivity ($k=N$) maximizes the speed of consensus.
This speed defines the response's cutoff frequency and, thus, an
increase in it improves the collective response to slow-changing environmental perturbations.
However, the short time-scale response of a decentralized system
(i.e. subjected to higher frequency components) cannot be inferred
solely from the large time-scale one corresponding to the consensus speed. In
that regime, the response is found to have a nontrivial relationship with the
degree of the interaction network. In fact, all the network models presented
here show an intrinsic trade-off in this respect: any change that increases
the low-frequency response decreases the high-frequency one, and vice versa.

The linear distributed consensus is a general model with applications in a wide range of social and artificial systems.
In general, these systems will have different design constraints in connectivity (arising, for example, from a spatial embedding of the nodes)
which make certain network models more relevant than others.
We performed two exploratory analyses that indicate that these results are relatively general.
On the one hand, when the ring network is allowed to have weighted links, the optimal distribution of weights is comparable to the distribution in the unweighted case.
On the other hand, when no structure is imposed, discrete optimization on small systems shows that the optimal degree distribution is still such that $\langle k\rangle^*$ decreases with $\omega$.

In practice, many multi-agent applications---such as artificial swarms---require an explicit or implicit design of the interaction network or interaction model.
This process usually involves numerous trade-offs owing to numerous design constraints~\cite{bouffanais15:_desig_contr_swarm_dynam}, and thus the network is designed to facilitate certain desired qualities of the system---such as robustness, scalability, and responsiveness.
What the findings presented here teach us is that, for multi-agent systems to be responsive, there is not one optimal interaction network that can guarantee responsiveness in general.
Instead, the connectivity can only be optimized to yield maximal responsiveness at a particular time-scale.
For all the cases considered here, an increase in responsiveness to low frequency signals implies a decrease in the response to high frequencies, and vice versa.

In order to showcase the applicability of our findings to the practical arena, we have performed a series of experiments with a set of 11 swarming robots that seek to emulate the behavior of a ``leader'' by performing a classical example of collective motion.
These robots communicate their state to a given set of neighbors and align their direction of motion to the average of said neighbors.
The experiments confirm that, when the leader changes its direction slowly, the agents are better at following the leader the more connected they are, and that the opposite is true for fast changes.

In this work, we have considered how the degree of the network affects the responsiveness of multi-agent systems performing distributed consensus.
However, it is not possible to change the degree distribution of a network without also changing its other properties such as clustering or shortest path.
While it is known that at high enough frequency the response is determined only by the mean degree~\cite{Mat17},
the question of which structural properties of the network are good predictors of a system's response at different regimes remains open.


\section*{Methods}

\subsection*{Distributed Linear Consensus}

Let us consider a group of $N+1$ identical agents performing a distributed consensus protocol on their scalar state-variable $x_i(t)$.
The dynamics of the system is determined by the state vector $\mathbf{X}(t) = \{x_i(t);i=0,\ldots,N\}$ and the adjacency matrix of the underlying graph ${\mathbf A}=\{a_{ij};i,j=0,\ldots,N\}$, where $a_{ij}=1$ if agent $i$ is connected to $j$ and 0 otherwise.
Given a certain connectivity graph, the system evolves according to
\begin{eqnarray}
  \frac{dx_i}{dt} &= \frac{\omega_0}{k_i}\sum_{j=0}^N a_{ij} \big( x_j(t) -
  x_i(t) \big) ,   \nonumber \\
  &= \sum_{j=0}^N w_{ij} x_j(t) ,
  \label{eq:local-consensus}
\end{eqnarray}
where $\omega_0$ is the natural response frequency of our identical agents, and $k_i = \sum_{j=0}^N a_{ij}$ is the degree (or number of neighbors) of agent $i$.
The quantity $w_{ij}=\omega_0 (a_{ij}/k_i-\delta_{ij})$---where $\delta_{ij}$ is a Kronecker delta---is introduced for the sake of a compact notation.
Note that, by definition, $w_{ii} = -\omega_0$ and $\sum_j w_{ij} = 0$ for all $i$.

To model the response of the system, we consider a leader-follower consensus scenario where one agent---say agent $i=0$, the ``leader''---does not abide by the dynamics of Eq.~\ref{eq:local-consensus}, but instead follows an arbitrary trajectory $x_0(t) = u(t)$.
In the presence of this single leader, Eq.~\ref{eq:local-consensus} can be recast as
\begin{equation}
  \frac{d x_i}{dt} = \sum_{j=1}^N w_{ij} x_j(t) + w_{i0} u(t) ,
  \label{eq:dynamics}
\end{equation}
for $i=1,\cdots,N$.
The solution of Eq.~\ref{eq:dynamics} (up to an integration constant) can be written compactly in matrix notation on the frequency domain as
\begin{equation}
    {\mathbf X}(\omega) =
    {(i\omega\mathbf{I}-\W)}^{-1}\WL u(\omega) , \
    \label{eq:matrix_solution}
\end{equation}
where $\mathbf{I}$ is the identity matrix of dimension $N$, $\W=\{w_{ij}\}$ is the $N\times N$ consensus protocol matrix between the follower agents (also known as state matrix ${\bf A}$ in LTI systems), and $\WL=\{w_{i0}\}$ is the $N\times 1$ consensus protocol matrix between the followers and the leader (also known as input matrix ${\bf B}$).

The response function or susceptibility measures the capacity of the multi-agent system to  follow the leader's trajectory, $u(t)$, and can be expressed in the frequency domain \cite{ogata10:_moder_contr_engin} as
\begin{equation}
    {\mathbf H}(\omega) =
  \left(\frac{\delta{\mathbf X}}{\delta u}\right)(\omega) =
  {(i\omega\mathbf{I}-\W)}^{-1}\WL.
  \label{eq:gain}
\end{equation}
The entries of the vector ${\mathbf H} = \{h_i\}_{i=1,\cdots,N}$ correspond to the frequency response of each individual agent, with $|h_{i}(\omega)|\le1$ for all $i$ and $\omega$~\cite{ogata10:_moder_contr_engin}.
As is clear from Eq.~\ref{eq:gain}, the response functions have a nontrivial dependency on the topology of the agents' connectivity through the entries of $\W$ and $\WL$.
The collective response of the system can be characterized by performing a singular value decomposition of $\mathbf{H}$, giving a single singular value $\sigma^2 = \sum_i |h_i|^2 = H^2$.
Throughout this work, we will use $H^2(\omega)$ (or its normalized form, $H^2/N$) as a measure of the collective frequency response of multi-agent systems.
From Eq.~\ref{eq:matrix_solution}, one can see that if $u(t)$ is taken to be a white-noise stochastic perturbation, the fluctuations in the collective will be correlated as $\langle \mathbf{X}^\dag\mathbf{X}\rangle \propto H^2$.

\subsection*{Models of Interaction Network}
\label{methods:networks}

To study the effect of the number of neighbors on the collective frequency response, we have considered a set of network topologies including one- and two-dimensional regular periodic lattices (i.e. a ring and a mesh), connected caveman model graphs~\cite{Watts99}, and regular random graphs.

We have chosen these models because they feature a fairly regular structure---minimizing the effect of the leader's location---and they allow for a fine control of the degree of the agents from $k=2$ or 4 up to an all-to-all connectivity ($k=N$).
They are also guaranteed to be connected, and thus have $H^2(\omega=0) = N$ for any $k$, with the exception of the random regular graph at low degrees ($k\lesssim\log{N}$)~\cite{bollobas1998random}.
All these models provide unweighted, undirected interaction networks without self-loops.

\begin{description}
    \item[Regular Lattice:] These networks are constructed by placing the agents in a regular square grid embedded in an Euclidean $d$-dimensional periodic space (a ring for $d=1$ and a torus for $d=2$) and connecting each node to its $k$ nearest neighbors on the grid.
    This structure guarantees that all nodes have the same centrality and degree, and we can set the leader arbitrarily as the first agent without loss of generality.

    For the 1D ring, the degree can be any even value between $2$ and $N$.
    For the 2D mesh, only multiples of 4 are valid degrees.

    \item[Caveman Model:]
The connected caveman graph is composed of $n$ complete subgraphs with $k+1$ nodes, where one edge from each cluster is modified so that it connects neighboring clusters~\cite{Watts99}.
Note that the number of nodes is $N=(k+1)n$ and the average degree is $k$ (there are $n$ nodes with a degree of $k+1$ and another $k$ with $k-1$, the rest have a degree of $k$).

We choose the number of agents $N$ to be a highly composite number (such as 720, 840, or 1260) in order to have a high resolution on the possible values of the degree $k$ while keeping $N$ constant.

    \item[Regular Random Networks:]
Finally, we consider the case of regular random graphs, i.e. graphs randomly sampled from all the possible $k$-regular graphs with $N$ nodes.
The graphs sampled are expected to be less structured than the lattices or caveman graphs, typically having significantly lower diameter and clustering coefficient.

The frequency response of random networks presented in Fig.~\ref{fig:optimal} is obtained by averaging over 100 samples for each frequency.
\end{description}

\subsection*{Weight Optimization}\label{methods:optimal}

The optimal distribution of weights $\mathbf{A}$ (or equivalently $\mathbf{W}$) for collective frequency response can be obtained by solving
\begin{equation}
    \frac{\delta H^2}{\delta \mathbf{W}} = 0 \, ,
\end{equation}
with the conditions that $\sum_{j=0}^N w_{ij} = 0$  for all $j$.
This gradient can be written as
\begin{eqnarray}
\frac{\partial H^2}{\partial w_{ij}} &=& {[\WL^\dag{(i\omega-\W)}^{-1\dag}{(i\omega-\W)}^{-2}]}_{i} {(\WL)}_{j} + \mathrm{h.c}\, , \nonumber \\
\frac{\partial H^2}{\partial w_{i0}} &=& {[\WL^\dag{(i\omega-\W)}^{-1\dag}{(i\omega-\W)}^{-1}]}_{i} + \mathrm{h.c}\, ,\nonumber \\
\frac{\partial H^2}{\partial w_{0j}} &=& 0 \, ,
\end{eqnarray}
where h.c.~stands for the Hermitian conjugate of the preceding expression.

If no constraints are imposed on the weights, a trivial solution is obtained in which the response is maximized by having all the agents connected only to the leader.
In order for this study to be applicable to cases where the leader is not known in advance and where it may even change with time, one needs to impose some additional symmetries either in $H^2$ or in $\mathbf{W}$ directly.

One option to introduce the symmetry in $H^2$ is to compute the frequency response not by fixing the leader to be a particular agent but to average over all the possible leaders instead.
This option is appropriate for small systems, but for relatively large systems it is more suitable to impose symmetries on $\mathbf{W}$ instead in order to reduce the number of variables (which otherwise grows as $N^2$).

To impose arbitrary symmetries in the system, one can constrain the weights with a given parametrization
\begin{equation}
    w_{ij} = F(i, j, \{c_k\}) \, ,
\end{equation}
where $\{c_k\}$ is a set of free parameters.
The gradient of $H^2$ with respect to these parameters is
\begin{eqnarray}
    & \frac{\partial H^2}{\partial c_k} = \sum_{ij} \frac{\partial H^2}{\partial w_{ij}} \frac{\partial w_{ij}}{\partial c_k}
    = \WL^\dag(i\omega-\W)^{-1\dag}(i\omega-\W)^{-2} \frac{\partial\W}{\partial c_k} \WL + \nonumber \\
    &+ \WL^\dag(i\omega-\W)^{-1\dag}(i\omega-\W)^{-1}\frac{\partial\WL}{\partial c_k} + \mathrm{h.c} \, .
\end{eqnarray}

We consider the case where the connection between agents depends only on the topological distance between them.
Given a measure of topological distance $d(i,j)$, this condition can be written as a linear parametrization of the form
\begin{equation}
    w_{ij} = \sum_k c_k m^{k}_{ij} \, ,
    \label{eq:linear_parametrization}
\end{equation}
where
\begin{equation}
    m^{k}_{ij} = \left\{
        \begin{array}{lr}
        1 & \mathrm{if}\quad d(i,j)=k \\
        0 & \mathrm{otherwise}
        \end{array}
    \right. \, .
\end{equation}

With a linear parametrization, we obtain a close form for the gradient as
\begin{eqnarray}
    & \frac{\partial H^2}{\partial c_k} =
    \WL^\dag(i\omega-\W)^{-1\dag}(i\omega-\W)^{-2} {\mathbf M}^k \WL + \nonumber \\
    &+ \WL^\dag(i\omega-\W)^{-1\dag}(i\omega-\W)^{-1}{\mathbf M}^k_L + \mathrm{h.c} \, ,
\end{eqnarray}
where ${\mathbf M}^k = \{m^k_{ij}\}$ and ${\mathbf M}^k_L = \{m^k_{i0}\}$.

The normalization of the weights during optimization can be imposed through a Lagrange multiplier.
Thus, instead of maximizing $H^2$, we define a cost function of the form
\begin{equation}
    {\mathcal L} = H^2 - \frac{\lambda}{2} \sum_{i=0}^N\left|\sum_{j=0}^N w_{ij}\right|^2 \, .
    \label{eq:cost}
\end{equation}
Using this cost function, the optimization problem over $w_{ij}$ with $i\neq j$ (the diagonal terms are fixed to $w_{ii}=-1$) can be written as
\begin{eqnarray}
    \frac{\partial {\mathcal L}}{\partial w_{ij}} &= \frac{\partial H^2}{\partial w_{ij}} - \lambda\sum_{l=0}^N w_{il} = 0
    \nonumber \\
    \frac{\partial {\mathcal L}}{\partial \lambda} &=
-\frac{1}{2}\sum_{ij}{(\mathbf{W}^\dag \mathbf{W})}_{ij} = 0 \, .
\label{optimal_problem}
\end{eqnarray}

For the linear parametrization of Eq.~\ref{eq:linear_parametrization}, the gradient of the cost function with respect to $c_k$ can be written as
\begin{equation}
    \frac{\partial \mathcal L}{\partial c_k} =
    \frac{\partial H^2}{\partial c_k} - \lambda\sum_{ij}(\mathbf{W}^T{\mathbf M^k})_{ij} \, .
\end{equation}

\subsection*{Robotic Platform}\label{methods:robot}

For the experimental validation, we have used a differential-drive robot developed in-house as a low-cost tool for robotics research~\cite{evobot}.
The platform is equipped with six infrared rangefinders, an inertial measurement unit (IMU), two wheel encoders, and two light sensors. %

\begin{figure}
    \begin{minipage}[l]{0.6\linewidth}
        \includegraphics[width=0.9\linewidth]{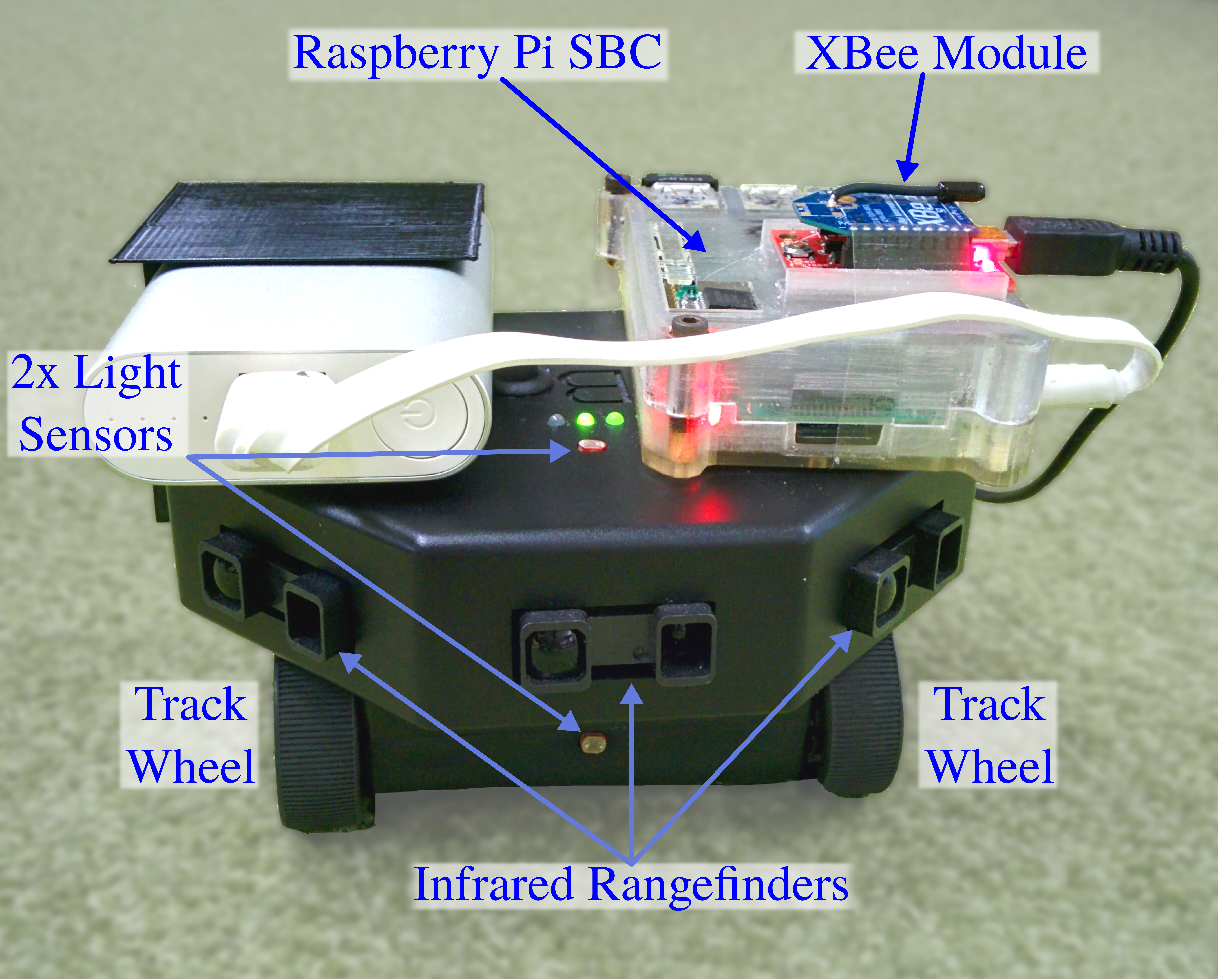}
    \end{minipage}
    \begin{minipage}[r]{0.4\linewidth}
        \caption{Robotic platform used in the experiments. The SBC and XBee module attached to the robot provides autonomy and distributed communications to the unit, making it able to swarm and perform decentralized collective motion such as heading consensus.}
        \label{fig:ebot}
\end{minipage}
\end{figure}

The robots are granted collective autonomy and distributed communication by attaching a ``swarm-enabling'' unit~\cite{chamanbaz17:_swarm_enabl_techn_multi_robot_system} composed of a single-board computer (SBC) and an XBee module~\cite{xbee}, see Fig.~\ref{fig:ebot}.
This unit interfaces with the robot via Bluetooth and is responsible for implementing the cooperative control algorithms, communicating with other robots, and controlling the motion of the robot.

The units communicate via radio signals sent over the distributed, dynamic mesh established by the XBee modules.
Since physical limitations impose a maximum range at which these signals can be sent, the swarm has a natural ``metric'' interaction model, meaning that each agent is able to communicate with any other agents within a given distance.
Field experiments with aquatic autonomous surface vehicles using a similar setup for communication~\cite{zoss17:_distr_system_auton_buoys_scalab} have shown that, when the number of agents is large, the interaction model is also weakly density-dependent, deviating from a pure metric model.
However, for the current experimental setup there is no practical spatial limitation on communication between the agents, and instead the different interaction networks are explored by tuning the cooperative control rule.

To control the robot, the SBC processes the data from the IMU and encoders through a K\'alm\'an filter to have an accurate estimation of the platform's location.
Then, a Proportional Integral Derivative (PID) controller allows it to adjust the trajectory of the robot.
The PID coefficients are tuned using the Ziegler--Nichols frequency response method~\cite{ogata10:_moder_contr_engin}.

The behavior of the robot, defined by the cooperative control algorithm, is implemented in the swarm-enabling unit using the in-house \texttt{marabunta} package~\cite{chamanbaz17:_swarm_enabl_techn_multi_robot_system}.
The software is designed prioritizing platform-agnostic development of cooperative control rules, portability, and a simple workflow to facilitate fast prototyping.
In the study of collective motion, these behavior rules typically take the form of simple, local, iterative algorithms where the velocity of an agent is defined in terms of its own state, the local environment, and the state of the agents in a certain neighborhood.

\subsection*{Heading Consensus Experiments}\label{methods:experiment}

To measure empirically the collective frequency response of a robotic swarm, we have performed leader-follower heading consensus~\cite{ref:ren} experiments using 11 robots.
One of them is designated as the ``leader'' and its behavior is a simple rotational motion at constant frequency $\omega$.
We label it as leader because its behavior does not depend on the rest of the swarm, but the agents have no way of distinguishing the leader from any other agent.
The other ten ``follower'' robots perform a heading consensus algorithm using the information received from a given set of neighboring agents, irrespectively of whether a neighbor is a leader or a follower.

The heading consensus algorithm determines a target heading $\bar{\theta}$ for each robot of the form
\begin{equation}
    \bar{\theta}_i = \langle \theta_j \rangle_{j\sim i} = \arctan\left(\frac{\sum_{j\sim i} \sin\theta_j}{\sum_{j\sim i} \cos\theta_j}\right)\, , \label{eq:heading_consensus}
\end{equation}
where $j\sim i$ denotes the neighbors of $i$ and $\langle\cdot\rangle$ an angular average.
Each follower robot updates its target heading asynchronously every $\Delta T = 0.1$~s.
The information of each neighbor's state is also updated every $0.1$ seconds, but not necessarily concurrently with each other or with the update of Eq.~\ref{eq:heading_consensus}.
The robots are interconnected with a ring network topology such that each individual unit can only communicate with $k$ other robots.

This heading consensus algorithm used in the experiments is a discrete-time equivalent to a linear consensus algorithm where the modulus of the state variable is fixed, i.e. it is equivalent to
\begin{equation}
    \frac{d\theta_i}{dt} = \omega_0 (\langle \theta_j \rangle_{j\sim i}-\theta_i)
\end{equation}
if $\omega_0\Delta T \gg 1$ and for times $t\gg \Delta T$.


Since the consensus algorithm is nonlinear, we cannot use the LTI framework to define the frequency response as Eq.~\ref{eq:gain}.
Instead, we use an equivalent response metric by defining the state of an agent as $x_i(t) = e^{-i\theta_i(t)}$, where $\theta_i$ is its heading.
The capacity of an agent $i$ to follow the leader $L$ is then measured by
\begin{equation}
    H_i = \Re\left[\frac{1}{T} \int_0^T \frac{x_i(t)}{x_L(t)} dt\right] =
    \frac{1}{T} \int_0^T
    \cos(\theta_i - \theta_L) dt \, ,
    \label{eq:experimental_performance}
\end{equation}
where $T$ is the duration of the experiment and $|H_i| \le 1$.
The collective frequency response of the system is then measured by
\begin{equation}
    H^2 = \sum_{i=1}^N |H_i|^2 ,
    \label{eq:experimental_gain}
\end{equation}
with $0 \le H^2 \le N$.

\section*{Acknowledgments}
\paragraph*{Funding}
This work was supported by a MOE-Tier 1 Grant \#SUTDT2017001, and a SUTD-MIT International Design Centre Grant IDG31700107.

\paragraph*{Competing Interests}
The authors declare no competing interests.
\paragraph*{Author Contributions}
D.M. and R.B. designed the study and the experiment.
D.M. and N.H. developed the analytical and numerical tools and analyzed the data.
V.H. and M.C. performed the experiments.
All authors wrote the manuscript.

\paragraph*{Data and materials availability}
All data needed to evaluate the conclusions in the paper are present in the paper and/or the Supplementary Materials. Additional data available from authors upon request.
\paragraph*{Supplementary Materials}
Movie M1:
Experiment to measure the collective response in leader-follower heading consensus of a swarm of $N + 1 = 11$ land robots with a low-frequency input signal.
Movie M2:
Same experiment performed with a high-frequency input signal.

\end{document}